\documentclass[12pt]{article}
\usepackage{amsmath}
\usepackage{amssymb}
\usepackage{amsfonts}

\renewcommand{\theequation}{\arabic{section}.\arabic{equation}}

\newcommand{\field}[1]{\mathbb{#1}}
\newcommand{\R}{\field{R}}
\newcommand{\N}{\field{N}}
\DeclareMathOperator{\sech}{sech}

\title{
Extending Romanovski polynomials in quantum mechanics}

\author{C. Quesne\thanks{Electronic address: cquesne@ulb.ac.be}\\
{\small\sl Physique Nucl\'eaire Th\'eorique et Physique Math\'ematique, 
Universit\'e Libre de Bruxelles,} \\ 
{\small \sl Campus de la Plaine CP229, Boulevard~du Triomphe, B-1050
Brussels, Belgium}}
\date{ }
\begin{document}
\baselineskip=22pt plus 1pt minus 1pt
\maketitle

\begin{abstract}
Some extensions of the (third-class) Romanovski polynomials (also called Romanovski/pseudo-Jacobi polynomials), which appear in bound-state wavefunctions of rationally-extended Scarf II and Rosen-Morse I potentials, are considered. For the former potentials, the generalized polynomials satisfy a finite orthogonality relation, while for the latter an infinite set of relations among polynomials with degree-dependent parameters is obtained. Both types of relations are counterparts of those known for conventional polynomials. In the absence of any direct information on the zeros of the Romanovski polynomials present in denominators, the regularity of the constructed potentials is checked by taking advantage of the disconjugacy properties of second-order differential equations of Schr\"odinger type. It is also shown that on going from Scarf I to Scarf II or from Rosen-Morse II to Rosen-Morse I potentials, the variety of rational extensions is narrowed down from types I, II, and III to type III only.  
\end{abstract}

\vspace{0.5cm}


 
\newpage
%
%
\section{INTRODUCTION}

Systems of orthogonal polynomials satisfying a hypergeometric-type equation appear in many problems of applied mathematics and mathematical physics, for instance in quantum mechanics. The better known are the infinite systems of Hermite, Laguerre, and Jacobi classical orthogonal polynomials \cite{nikiforov}. Apart from these, there, however, exist three less known finite systems of orthogonal polynomials, which were discovered in 1884 by Routh \cite{routh} as polynomials defined in the complex plane and rediscovered in 1929 by Romanovski \cite{romanovski} as real polynomials. The three finite classes are known as Romanovski/Jacobi, Romanovski/Bessel, and Romanovski/pseudo-Jacobi \cite{lesky}, although the name Romanovski is most often used only for the last one \cite{raposo}. In quantum mechanics, they make their appearance in bound-state wavefunctions of several well-known shape-invariant potentials \cite{cotfas04, cotfas06}. The three categories of Romanovski polynomials are directly connected with generalized P\"oschl-Teller, Morse, and Scarf II potentials, respectively, and in a more indirect way to several others.\par
%
%
During the last few years, on the other hand, much work has been devoted to the construction of so-called exceptional orthogonal polynomials (EOP), which are new complete and orthogonal non-hypergeometric-type polynomial systems extending the classical families of Hermite, Laguerre, and Jacobi (see, e.g., Refs.~\cite{gomez04, gomez09, gomez10, gomez12a, gomez12b, gomez12c, cq08, bagchi, cq09, cq11a, cq11b, cq12a, cq12b, odake09, odake10, sasaki, odake11, odake13a, odake13b, grandati11a, grandati11b, grandati12a, grandati12b, grandati13, ho11a, ho11b, fellows, marquette, gomez13} and references quoted therein). In contrast with the latter, the former admit some gaps in the sequence of their degrees, the total number of them being referred to as the codimension. The EOP turn out to be basic ingredients of bound-state wavefunctions  for rationally-extended shape-invariant potentials.\par
%
%
Some of the EOP families considered so far are, however, made of only a finite number of polynomials and may not therefore be considered as extensions of the classical families. A clear classification similar to that of hypergeometric-type polynomials is indeed still lacking. As a first step in this direction, we plan to construct in this paper some extensions of the third-class Romanovski polynomials that appear in rational extensions of Scarf II and Rosen-Morse I potentials.\par
%
%
After presenting the classification of hypergeometric-type orthogonal polynomials in Sec.~II, we review the use of Romanovski polynomials in the bound-state problem for Scarf II and Rosen-Morse I potentials in Sec.~III. Section IV  then deals with the construction of rational extensions of these potentials by using a standard supersymmetric quantum mechanical (SUSYQM) approach \cite{cooper}. This will lead to corresponding extensions of Romanovski polynomials. Finally, Section V contains the conclusion.\par
%
%
\section{ORTHOGONAL POLYNOMIALS OF HYPERGEOMETRIC TYPE}

Equations of hypergeometric type are second-order differential equations
\begin{equation}
  \left(\sigma(z) \frac{d^2}{dz^2} + \tau(z) \frac{d}{dz} + \lambda\right) F(z) = 0,  \label{eq:hyper}
\end{equation}
where $\sigma(z)$ and $\tau(z)$ are polynomials of at most second and first degree, respectively, and $\lambda$ is a constant \cite{nikiforov}. Each of them can be reduced to the self-adjoint form
\begin{equation}
  \frac{d}{dz} \left(\sigma(z) \rho(z) \frac{d}{dz} F(z)\right) + \lambda \rho(z) F(z) = 0 
\end{equation}
by choosing a function $\rho(z)$ satisfying Pearson's differential equation
\begin{equation}
  \frac{d}{dz} [\sigma(z) \rho(z)] = \tau(z) \rho(z).  \label{eq:pearson}
\end{equation}
\par
%
%
Equation (\ref{eq:hyper}) is usually considered on an interval $(a,b)$, such that
\begin{equation}
\begin{split}
  & \sigma(z) > 0 \qquad \text{for all $z \in (a,b)$}, \\
  & \rho(z) > 0 \qquad \text{for all $z \in (a,b)$}, \\
  & \lim_{z \to a} \sigma(z) \rho(z) = \lim_{z \to b} \sigma(z) \rho(z) = 0. 
\end{split}
\end{equation}
Since the form of Eq.~(\ref{eq:hyper}) remains invariant under a change of variable $z \to cz + d$, it is sufficient to analyse the cases presented in Table I, where the restrictions on parameters $\alpha$ and $\beta$ are imposed by the existence of the interval $(a,b)$ \cite{cotfas04, cotfas06}.\par
%
%
\begin{table}[h!]

\caption{The main particular cases of Eq.~(\ref{eq:hyper}).}

\begin{center}
\begin{tabular}{lllll}
  \hline\hline\\[-0.2cm]
  $\sigma(z)$ & $\tau(z)$ & $\rho(z)$ & $\alpha$, $\beta$ & $(a,b)$\\[0.2cm]
  \hline\\[-0.2cm]
  1 & $\alpha z + \beta$ & $e^{\frac{1}{2}\alpha z^2 + \beta z}$ & $\alpha < 0$ & $\R$ \\[0.2cm]
  $z$ & $\alpha z + \beta$ & $z^{\beta-1} e^{\alpha z}$ & $\alpha < 0$, $\beta > 0$ & $(0, \infty)$ \\[0.2cm]
  $1 - z^2$ & $\alpha z + \beta$ & $(1+z)^{- \frac{1}{2}(\alpha-\beta)-1} (1-z)^{- \frac{1}{2}(\alpha+\beta)-1}$
        & $\alpha < \beta < - \alpha$ & $(-1,1)$ \\[0.2cm]
  $z^2 - 1$ & $\alpha z + \beta$ & $(z+1)^{\frac{1}{2}(\alpha-\beta)-1} (z-1)^{\frac{1}{2}(\alpha+\beta)-1}$
        & $-\beta < \alpha < 0$ & $(1, \infty)$ \\[0.2cm]
  $z^2$ & $\alpha z + \beta$ & $z^{\alpha-2} e^{- \frac{\beta}{z}}$ & $\alpha < 0$, $\beta > 0$ & $(0, \infty)$
        \\[0.2cm]
  $z^2 + 1$ & $\alpha z + \beta$ & $(1 + z^2)^{\frac{\alpha}{2} - 1} e^{\beta \arctan z}$ & $\alpha < 0$ & 
        $\R$ \\[0.2cm]
  \hline \hline
\end{tabular}
\end{center}

\end{table}
\par
%
%
It is well known \cite{nikiforov} that for  $\lambda = \lambda_{\nu}$, where
\begin{equation}
  \lambda_{\nu} = - \frac{1}{2} \nu(\nu-1) \frac{d^2}{dz^2} \sigma(z) - \nu \frac{d}{dz} \tau(z), \qquad \nu
  \in \N, 
\end{equation}
Eq.~(\ref{eq:hyper}) admits a polynomial solution $F_{\nu}(z)$ of at most $\nu$th degree,
\begin{equation}
  \left(\sigma(z) \frac{d^2}{dz^2} + \tau(z) \frac{d}{dz} + \lambda_{\nu}\right) F_{\nu}(z) = 0.
\end{equation}
If the degree of $F_{\nu}(z)$ is $\nu$, then it satisfies the Rodrigues formula
\begin{equation}
  F_{\nu}(z) = \frac{B_{\nu}}{\rho(z)} \frac{d^{\nu}}{dz^{\nu}} [\sigma^{\nu}(z) \rho(z)],  \label{eq:rodrigues}
\end{equation}
where $B_{\nu}$ is a constant depending on the normalization chosen for $F_{\nu}(z)$. It is possible to show \cite{cotfas04, cotfas06} that the set of polynomials $\{F_{\nu}(z) \mid \nu < \bar{\nu}\}$, with $\bar{\nu}$ defined by
\begin{equation}
  \bar{\nu} = \begin{cases}
    \infty & \text{for $\sigma(z) \in \{1, z, 1-z^2\}$},\\
    \frac{1-\alpha}{2} & \text{for $\sigma(z) \in \{z^2-1, z^2, z^2+1\}$,} 
  \end{cases}
\end{equation}
satisfies the following properties:
\par
\begin{itemize}
  \item[a)] $\{F_{\nu}(z) \mid \nu < \bar{\nu}\}$ is a system of orthogonal polynomials with weight function 
     $\rho(z)$ in $(a,b)$. 
  \item[b)] $F_{\nu}(z)$ is a polynomial of degree $\nu$ for any $\nu < \bar{\nu}$. 
  \item[c)] The function $F_{\nu}(z) \sqrt{\rho(z)}$ is square integrable on $(a,b)$ for any $\nu < \bar{\nu}$. 
  \item[d)] A three-term recursion relation
     \begin{equation}
       z F_{\nu}(z) = \alpha_{\nu} F_{\nu+1}(z) + \beta_{\nu} F_{\nu}(z) + \gamma_{\nu} F_{\nu-1}(z)
       \label{eq:recursion}
     \end{equation}
     is satisfied for $1 < \nu+1 < \bar{\nu}$.
  \item[e)] The zeros of $F_{\nu}(z)$ are simple and lie in the interval $(a,b)$ for any $\nu < \bar{\nu}$.
\end{itemize}
\par
%
%
{}For the first three cases of Table~I, corresponding to $\bar{\nu} = \infty$, we recover well-known properties of the three families of Hermite, Laguerre, and Jacobi classical orthogonal polynomials. The last three cases of Table~I only differ from them by the replacement of $\bar{\nu} = \infty$ by $\bar{\nu} = (1-\alpha)/2$ and, consequently, by that of infinite sets by finite ones. They correspond to the three families introduced by Romanovski, whose members may be alternatively expressed in terms of Jacobi polynomials outside of the interval $-1 < z < 1$ on which they are usually defined, of generalized Bessel polynomials \cite{krall} (which can be written in terms of Laguerre polynomials) and of pseudo-Jacobi polynomials (i.e., Jacobi polynomials with complex variable and parameters), respectively (see Refs.~\cite{cotfas04, cotfas06} for detailed expressions).\par
%
%
In the next Section, we consider in more detail the last ones and some of their occurrences in quantum mechanics.\par
%
%
\section{ROMANOVSKI POLYNOMIALS IN THE SCARF II AND ROSEN-MORSE I POTENTIAL PROBLEMS}

\setcounter{equation}{0}

\subsection{Romanovski polynomials}

To conform with the notations used in \cite{raposo}, let us change $\alpha$ and $\beta$ in the last row of Table~I into $2\beta$ and $\alpha$, respectively. Then we get the so-called Romanovski polynomials $R^{(\alpha, \beta)}_{\nu}(z)$, which are solutions of the differential equation
\begin{equation}
  \left((1+z^2) \frac{d^2}{dz^2} + (2\beta z + \alpha) \frac{d}{dz} - \nu (\nu-1+2\beta)\right)
  R^{(\alpha, \beta)}_{\nu}(z) = 0, \quad \beta < 0, \quad -\infty < z < \infty.  \label{eq:eq-diff-R}
\end{equation}
The corresponding weight function is given by
\begin{equation}
  \rho(z) = (1+z^2)^{\beta-1} e^{\alpha \arctan z}.
\end{equation}
We choose to normalize the polynomials in such a way that the constant $B_{\nu}$ in Rodrigues formula (\ref{eq:rodrigues}) is given by $B_{\nu} = (2^{\nu} \nu!)^{-1}$, which means that their highest-degree term is $2^{-\nu} \binom{2\nu+2\beta-2}{\nu} z^{\nu}$.\par
%
%
The orthogonality relation satisfied by the polynomials can be written as
\begin{equation}
  \int_{-\infty}^{\infty} dz\, (1+z^2)^{\beta-1} e^{\alpha \arctan z} R^{(\alpha, \beta)}_{\nu'}(z) 
  R^{(\alpha, \beta)}_{\nu}(z) = 0 \qquad \text{if $\nu' \ne \nu$ and $\nu, \nu' < \frac{1}{2} - \beta$},
  \label{eq:R-ortho}
\end{equation}
while the normalization integral
\begin{equation}
  \int_{-\infty}^{\infty} dz\, (1+z^2)^{\beta-1} e^{\alpha \arctan z} \left[R^{(\alpha, \beta)}_{\nu}(z)\right]^2
\end{equation}
converges for any $\nu < \frac{1}{2} - \beta$.\par
%
%
At this point, it is worth observing that a real orthogonalizing weight distribution has been proposed for the infinite set of Romanovski polynomials, termed twisted Jacobi polynomials, by solving a non-homogeneous version of Eq.~(\ref{eq:pearson}) \cite{kwon}. Such a weight, however, is neither explicit nor positive-definite.\par
%
%
{}From both the differential equation (\ref{eq:eq-diff-R}) and the Rodrigues formula (\ref{eq:rodrigues}), it is easy to see that the Romanovski polynomials can be expressed as Jacobi polynomials with complex variable and parameters in the form
\begin{equation}
  R^{(\alpha, \beta)}_{\nu}(z) = (- {\rm i})^{\nu} P^{(\alpha', \beta')}_{\nu}({\rm i} z), \label{eq:R-P}
\end{equation}
where
\begin{equation}
  \alpha' = \beta - 1 + {\rm i} \frac{\alpha}{2}, \qquad \beta' = \beta - 1 - {\rm i} \frac{\alpha}{2}, \qquad 
  \alpha = - {\rm i}(\alpha' - \beta'), \qquad \beta = \frac{1}{2}(\alpha' + \beta' + 2). \label{eq:R-P-parameters}
\end{equation}
It follows that the coefficients $\alpha_{\nu}$, $\beta_{\nu}$, $\gamma_{\nu}$ in the three-term recursion relation (\ref{eq:recursion}) are given by
\begin{equation}
\begin{split}
  & \alpha_{\nu}  = \frac{2(\nu+1)(\nu+2\beta-1)}{(2\nu+2\beta)(2\nu+2\beta-1)}, \\
  & \beta_{\nu}  = - \frac{\alpha(2\beta-2)}{(2\nu+2\beta)(2\nu+2\beta-2)}, \\ 
  & \gamma_{\nu}  = - \frac{2(\nu+\beta-1+{\rm i}\frac{\alpha}{2})(\nu+\beta-1-{\rm i}\frac{\alpha}{2})}
        {(2\nu+2\beta-1)(2\nu+2\beta-2)}.
\end{split}
\end{equation}
\par
%
%
It is also worth mentioning the property
\begin{equation}
  (1+z^2) \frac{d}{dz} R^{(\alpha,\beta)}_{\nu}(z) = \frac{\nu+2\beta-1}{2\nu+2\beta} \left\{- [\alpha +
  (2\nu+2\beta) z] R^{(\alpha,\beta)}_{\nu}(z) + 2(\nu+1) R^{(\alpha,\beta)}_{\nu+1}(z)\right\},
  \label{eq:diff-R}
\end{equation}
directly obtainable from the Rodrigues formula (\ref{eq:rodrigues}), as well as its variant
\begin{equation}
\begin{split}
  & (1+z^2) \frac{d}{dz} R^{(\alpha,\beta)}_{\nu}(z) = \frac{1}{2\nu+2\beta-2} \Bigl\{\nu [- \alpha + 
       (2\nu+2\beta-2) z] R^{(\alpha,\beta)}_{\nu}(z) \\ 
  & \quad + 2 \Bigl(\nu+\beta-1+{\rm i}\frac{\alpha}{2}\Bigr) \Bigl(\nu+\beta-1-{\rm i}\frac{\alpha}{2}\Bigr) 
       R^{(\alpha,\beta)}_{\nu-1}(z))\Bigr\}, \label{eq:diff-R-bis}
\end{split}
\end{equation}
resulting from the combination of Eqs.~(\ref{eq:recursion}) and (\ref{eq:diff-R}).\par
%
%
\subsection{Scarf II potential problem}

Let us now turn ourselves to the Schr\"odinger equation for the Scarf II potential \cite{cooper},
\begin{equation}
  \left(- \frac{d^2}{dx^2} + V_{A,B}(x)\right) \phi(x) = E \phi(x),  \label{eq:schrodinger}
\end{equation}
where 
\begin{equation}
  V_{A,B}(x) = [B^2 - A(A+1)] \sech^2 x + B (2A+1) \sech x \tanh x, \quad -\infty < x < \infty, \quad A>0.
  \label{eq:scarf-pot}
\end{equation}
It is known to lead to a finite number of bound states, whose energy and wavefunction are given by
\begin{equation}
  E^{(A)}_{\nu} = - (A-\nu)^2, \qquad \nu = 0, 1, \ldots, \nu_{\rm max}, \qquad A-1 \le \nu_{\rm max} < A,
\end{equation}
and
\begin{equation}
\begin{split}
  \psi_{\nu}^{(A,B)}(x) & \propto (\sech x)^A e^{- B \arctan(\sinh x)} 
        R_{\nu}^{(-2B, -A+\frac{1}{2})}(\sinh x) \\
  & \propto (-{\rm i})^{\nu} (\sech x)^A e^{- B \arctan(\sinh x)} P_{\nu}^{(-A-\frac{1}{2}-{\rm i}B, 
         -A-\frac{1}{2}+{\rm i}B)}({\rm i} \sinh x), 
\end{split}  \label{eq:scarf-wf}
\end{equation}
respectively.\par
%
%
In terms of the variable $z = \sinh x$ ($-\infty < z < \infty$), the Schr\"odinger equation (\ref{eq:schrodinger}) and the wavefunctions (\ref{eq:scarf-wf}) can be rewritten as
\begin{equation}
  \left(- (1+z^2) \frac{d^2}{dz^2} - z \frac{d}{dz} + \frac{B^2 - A(A+1)}{1+z^2} + B(2A+1) \frac{z}{1+z^2}
  - E \right) \phi(x(z)) = 0  \label{eq:scarf-eq}
\end{equation}
and
\begin{equation}
\begin{split}
  \psi_{\nu}^{(A,B)}(x) & \propto (z+{\rm i})^{-\frac{1}{2}(A+{\rm i}B)} (z-{\rm i})^{-\frac{1}{2}(A-{\rm i}B)}
        R_{\nu}^{(-2B, -A+\frac{1}{2})}(z) \\
  & \propto (-{\rm i})^{\nu} (z+{\rm i})^{-\frac{1}{2}(A+{\rm i}B)} (z-{\rm i})^{-\frac{1}{2}(A-{\rm i}B)} 
        P_{\nu}^{(-A-\frac{1}{2}-{\rm i}B, -A-\frac{1}{2}+{\rm i}B)}({\rm i}z). 
\end{split}  \label{eq:scarf-wf-bis} 
\end{equation}
From this, it follows that
\begin{equation}
\begin{split}
  \delta_{\nu',\nu} & = \int_{-\infty}^{+\infty} dx\, \psi_{\nu'}^{(A,B)}(x) \psi_{\nu}^{(A,B)}(x) \\
  & \propto \int_{-\infty}^{+\infty} dz\, (1+z^2)^{\beta-1} e^{\alpha \arctan z} R_{\nu'}^{(\alpha,\beta)}(z)
       R_{\nu}^{(\alpha,\beta)}(z)
\end{split}
\end{equation}
with $\nu, \nu' < \frac{1}{2} - \beta$, $\alpha \equiv -2B$, and $\beta \equiv -A+\frac{1}{2}$. Hence the finite number of orthonormal bound-state wavefunctions $\psi_{\nu}^{(A,B)}(x)$ of the Scarf II potential is directly connected with the finite number of corresponding Romanovski polynomials $R_{\nu}^{(\alpha,\beta)}(z)$ that are orthogonal and normalizable in the framework defined in Sec.~II (see Eq.~(\ref{eq:R-ortho})) \cite{raposo}.\par
%
%
{}For future use, it is worth observing that the Scarf II Schr\"odinger equation can be obtained by formally complexifying that of the Scarf I potential
\begin{equation}
  \tilde{V}_{A,B}(x) = [A(A-1) + B^2] \sec^2 x - B(2A-1) \sec x \tan x, \quad - \frac{\pi}{2} < x < \frac{\pi}{2},
  \quad 0 < B < A-1,
\end{equation}
which is characterized by an infinite number of bound states of energy
\begin{equation}
  \tilde{E}^{(A)}_{\nu} = (A+\nu)^2, \qquad \nu=0, 1, 2, \ldots,  \label{eq:scarf-E}
\end{equation}
and wavefunction
\begin{equation}
  \tilde{\psi}^{(A,B)}_{\nu}(x) \propto (1 - \sin x)^{\frac{1}{2}(A-B)} (1 + \sin x)^{\frac{1}{2}(A+B)} 
  P_{\nu}^{(A-B-\frac{1}{2}, A+B-\frac{1}{2})}(\sin x).
\end{equation}
In terms of the variable $w = \sin x$ ($-1 < w < 1$), the Scarf I Schr\"odinger equation can indeed be written as
\begin{equation}
  \left(- (1-w^2) \frac{d^2}{dw^2} + w \frac{d}{dw} + \frac{A(A-1)+B^2}{1-w^2} - B(2A-1) \frac{w}{1-w^2}
  - \tilde{E}\right) \tilde{\phi}(x(w)) = 0  \label{eq:scarf-I-eq}
\end{equation}
and the substitutions
\begin{equation}
  w = \sin x \to {\rm i}z = {\rm i}\sinh x, \qquad A \to -A, \qquad B \to {\rm i} B, \qquad \tilde{E} \to -E, \qquad
  \tilde{\phi} \to \phi  \label{eq:scarf-formal}
\end{equation}
transform it into Eq.~(\ref{eq:scarf-eq}). In such a transition, $\tilde{E}^{(A)}_{\nu}$ and $\tilde{\psi}^{(A,B)}_{\nu}(x)$ are changed into $E^{(A)}_{\nu}$ and $\psi_{\nu}^{(A,B)}(x)$, respectively. Note, however, that the range of the variables, the parameters, and the quantum numbers is completely modified.\par
%
%
\subsection{Rosen-Morse I potential problem}

Instead of (\ref{eq:scarf-pot}), let us consider the Rosen-Morse I potential \cite{cooper}
\begin{equation}
  V_{A,B}(x) = A(A-1) \csc^2 x + 2B \cot x, \qquad 0 < x < \pi, \qquad A \ge \frac{3}{2},  \label{eq:RM-I}
\end{equation}
in Eq.~(\ref{eq:schrodinger}). This potential has an infinite number of bound states, characterized by the energies
\begin{equation}
  E^{(A,B)}_{\nu} = (A+\nu)^2 - \frac{B^2}{(A+\nu)^2}, \qquad \nu=0, 1, 2, \ldots,  \label{eq:RM-I-E} 
\end{equation}
and the wavefunctions
\begin{equation}
\begin{split}
  \psi^{(A,B)}_{\nu}(x) & \propto (\sin x)^{A+\nu} \exp\left(\frac{B}{A+\nu}x\right) 
       R_{\nu}^{\left(-\frac{2B}{A+\nu}, -A-\nu+1\right)}(\cot x) \\
  & \propto (-{\rm i})^{\nu} (\sin x)^{A+\nu} \exp\left(\frac{B}{A+\nu}x\right)
       P_{\nu}^{\left(-A-\nu-\frac{{\rm i}B}{A+\nu}, -A-\nu+\frac{{\rm i}B}{A+\nu}\right)}({\rm i}\cot x). 
\end{split} \label{eq:RM-I-wf}
\end{equation}
\par
%
%
In terms of the variable $z = \cot x$ ($-\infty < z < \infty$), the Schr\"odinger equation (\ref{eq:schrodinger}) with potential (\ref{eq:RM-I}) can be rewritten as
\begin{equation}
  \left(- (1+z^2)^2 \frac{d^2}{dz^2} - 2z(1+z^2) \frac{d}{dz} + A(A-1)(1+z^2) + 2Bz - E\right) \phi(x(z)) = 0
  \label{eq:RM-I-eq}
\end{equation}
and the wavefunctions (\ref{eq:RM-I-wf}) become
\begin{equation}
\begin{split}
  \psi^{(A,B)}_{\nu}(x) & \propto (z+{\rm i})^{-\frac{1}{2}\left(A+\nu+\frac{{\rm i}B}{A+\nu}\right)}
       (z-{\rm i})^{-\frac{1}{2}\left(A+\nu-\frac{{\rm i}B}{A+\nu}\right)} 
       R_{\nu}^{\left(-\frac{2B}{A+\nu}, -A-\nu+1\right)}(z) \\
  & \propto (-{\rm i})^{\nu} (z+{\rm i})^{-\frac{1}{2}\left(A+\nu+\frac{{\rm i}B}{A+\nu}\right)}
       (z-{\rm i})^{-\frac{1}{2}\left(A+\nu-\frac{{\rm i}B}{A+\nu}\right)} \\
  & \quad \times P_{\nu}^{\left(-A-\nu-\frac{{\rm i}B}{A+\nu}, -A-\nu+\frac{{\rm i}B}{A+\nu}\right)}({\rm i}z).
\end{split} \label{eq:RM-I-wf-bis} 
\end{equation}
Here we note an important difference with respect to the wavefunctions (\ref{eq:scarf-wf-bis}) of the previous potential: the parameters of the Romanowski polynomials now not only depend on the potential parameters $A$, $B$, but also on the polynomial degree $\nu$. For different $\nu$ values, such polynomials do not therefore belong to the same (finite) family. As a consequence, the orthonormality relation for the infinite set of bound-state wavefunctions provides an infinite set of relations among Romanovski polynomials with parameters attached to the degree \cite{raposo},
\begin{equation}
\begin{split}
  \delta_{\nu',\nu} & = \int_0^{\pi} dx\, \psi_{\nu'}^{(A,B)}(x) \psi_{\nu}^{(A,B)}(x) \\
  & \propto \int_{-\infty}^{+\infty} dz\, (1+z^2)^{\frac{1}{2}(\beta_{\nu}+\beta_{\nu'})-2} 
       e^{\frac{1}{2}(\alpha_{\nu}+\alpha_{\nu'}) \arctan z} R_{\nu'}^{(\alpha_{\nu'},\beta_{\nu'})}(z)
       R_{\nu}^{(\alpha_{\nu},\beta_{\nu})}(z)
\end{split}  \label{eq:infinite-ortho}
\end{equation}
with $\nu$, $\nu'=0$, 1, 2, \ldots, $\alpha_{\nu} \equiv -2B/(A+\nu)$, $\alpha_{\nu'} \equiv -2B/(A+\nu')$, $\beta_{\nu} \equiv -A-\nu-1$, and $\beta_{\nu'} \equiv -A-\nu'-1$.\par
%
%
Here it is worth noting a useful relation between Eq.~(\ref{eq:RM-I-eq}) and the Schr\"odinger equation for the Rosen-Morse II potential
\begin{equation}
  \tilde{V}_{A,B}(x) = - A(A+1) \sech^2 x + 2B \tanh x, \quad -\infty < x < \infty, \quad A>0, \quad 0 < B < A^2.
\end{equation}
In terms of the variable $w = \tanh x$ ($-1 < w < 1$), the latter may indeed be written as
\begin{equation}
  \left(- (1-w^2)^2 \frac{d^2}{dw^2} + 2w(1-w^2) \frac{d}{dw} - A(A+1)(1-w^2) + 2Bw - \tilde{E}\right)
  \tilde{\phi}(x(w)) = 0.
\end{equation}
On applying the formal transformation
\begin{equation}
  w = \tanh x \to {\rm i}z = {\rm i} \cot x, \qquad A \to -A, \qquad B \to {\rm i}B, \qquad \tilde{E} \to - E,
  \qquad \tilde{\phi} \to \phi,  \label{eq:RM-formal}
\end{equation}
to the latter, we get Eq.~(\ref{eq:RM-I-eq}). In such a process, the bound-state energies and wavefunctions (\ref{eq:RM-I-E}) and (\ref{eq:RM-I-wf}) result from those of the Rosen-Morse II potential, given in Sec.~2.1 of Ref.~\cite{cq12b}, up to the range of all involved quantities.\par
%
%
\section{RATIONALLY-EXTENDED SCARF II AND ROSEN-MORSE I POTENTIALS AND CORRESPONDING ORTHOGONAL POLYNOMIALS}

\setcounter{equation}{0}

\subsection{Construction method}

In this Section, we plan to construct rational extensions of the potentials (\ref{eq:scarf-pot}) and (\ref{eq:RM-I}) and to determine the counterparts of the Romanovski polynomials appearing in the bound-state wavefunctions (\ref{eq:scarf-wf}) and (\ref{eq:RM-I-wf}).\par
%
%
{}For such a purpose, we make use of a first-order SUSYQM approach \cite{cooper}, wherein a pair of SUSY partner Hamiltonians
\begin{equation}
  H^{(\pm)} = - \frac{d^2}{dx^2} + V^{(\pm)}(x) - E
\end{equation}
is built with
\begin{equation}
  V^{(+)}(x) = V_{A',B}(x), \qquad V^{(-)}(x) = V_{A,B}(x) + V_{A,B,{\rm rat}}(x). \label{eq:partner-pot}
\end{equation}
Here $V^{(+)}(x)$ is a conventional potential with some translated parameter $A'$, while $V^{(-)}(x)$ contains an additional rational part $V_{A,B,{\rm rat}}(x)$. The construction uses a superpotential $W(x) = - \bigl(\log \phi(x)\bigr)'$, obtained from some nodeless seed solution $\phi(x)$ of the initial Schr\"odinger equation
\begin{equation}
  \left(- \frac{d^2}{dx^2} + V^{(+)}(x)\right) \phi(x) = E \phi(x)
\end{equation}
with energy $E$ (called factorization energy) below the ground-state energy of $V^{(+)}(x)$. From $W(x)$, we obtain the partner potentials in the form
\begin{equation}
  V^{(\pm)}(x) = W^2(x) \mp W'(x) + E,  \label{eq:partner-pot-W}
\end{equation}
while the partner Hamiltonians can be factorized as
\begin{equation}
\begin{split}
  & H^{(+)} = \hat{A}^{\dagger} \hat{A}, \qquad H^{(-)} = \hat{A} \hat{A}^{\dagger}, \\
  & \hat{A}^{\dagger} = - \frac{d}{dx} + W(x), \qquad \hat{A} = \frac{d}{dx} + W(x), \label{eq:partner-factor}
\end{split}
\end{equation}
and intertwine with $\hat{A}$ and $\hat{A}^{\dagger}$ as $\hat{A} H^{(+)} = H^{(-)} \hat{A}$, $\hat{A}^{\dagger} H^{(-)} = H^{(+)} \hat{A}^{\dagger}$. According to whether the inverse $\phi^{-1}(x)$ of the factorization function is normalizable or not, the partner Hamiltonian $H^{(-)}$ has an extra bound state of energy $E$ below the ground state of $H^{(+)}$ or it has the same spectrum as $H^{(+)}$.\par
%
%
To get a rationally-extended partner potential such as $V^{(-)}(x)$ in (\ref{eq:partner-pot}), we have to start from a factorization function of polynomial type. Since the polynomial-type solutions $\tilde{\phi}(x)$ of the Scarf I and Rosen-Morse II Schr\"odinger equations are well known, we will use the formal transformations (\ref{eq:scarf-formal}) and (\ref{eq:RM-formal}) to obtain those for the Scarf II and Rosen-Morse I equations, respectively. It will then remain to impose that the corresponding $E$ is smaller than the ground-state energy of $V^{(+)}(x)$, which in general implies some restrictions on the parameters, and to check that the resulting $\phi(x)$ is nodeless on the defining interval of $x$. To solve the last problem, the easiest way consists in making use of the disconjugacy properties of the Schr\"odinger equation for eigenvalues below the ground state \cite{hartman, coppel, bocher}. As a consequence of such properties, the nodeless character of the polynomial present in the denominator of $V^{(-)}(x)$ can be inferred from the equality of the signs it takes at both ends of the defining interval \cite{grandati12a, grandati13}.\par
%
%
\subsection{Rational extensions of the Scarf II potential and corresponding orthogonal polynomials}

Rational extensions of the Scarf I potential (or, equivalently, of the trigonometric P\"oschl-Teller potential) have been dealt with in many papers (see, e.g., Refs.~\cite{cq08, cq09, odake09, sasaki, odake11}). In Appendix A, we summarize (in the present notations) the main results for the polynomial-type solutions of the conventional Scarf I potential that are useful for deriving those of the Scarf II.\par
%
%
On performing substitutions (\ref{eq:scarf-formal}) in Eq.~(\ref{eq:A1}), we obtain the following four polynomial-type solutions of the Scarf II Schr\"odinger equation, given in Eqs.~(\ref{eq:schrodinger}) and (\ref{eq:scarf-pot}) or (\ref{eq:scarf-eq}),
\begin{equation}
\begin{split}
  & \phi_1(x) \propto (-{\rm i})^m (1-{\rm i}z)^{-\frac{1}{2}({\rm i}B+A)} (1+{\rm i}z)^{-\frac{1}{2}({\rm i}B-A-1)}
       P_m^{(-{\rm i}B-A-\frac{1}{2}, -{\rm i}B+A+\frac{1}{2})}({\rm i}z), \\ 
  & \phi_2(x) \propto (-{\rm i})^m (1-{\rm i}z)^{\frac{1}{2}({\rm i}B+A+1)} (1+{\rm i}z)^{\frac{1}{2}({\rm i}B-A)}
       P_m^{({\rm i}B+A+\frac{1}{2}, {\rm i}B-A-\frac{1}{2})}({\rm i}z), \\ 
  & \phi_3(x) \propto (-{\rm i})^m (1-{\rm i}z)^{\frac{1}{2}({\rm i}B+A+1)} (1+{\rm i}z)^{-\frac{1}{2}({\rm i}B-A-1)}
       P_m^{({\rm i}B+A+\frac{1}{2}, -{\rm i}B+A+\frac{1}{2})}({\rm i}z), \\ 
  & \phi_4(x) \propto (-{\rm i})^m (1-{\rm i}z)^{-\frac{1}{2}({\rm i}B+A)} (1+{\rm i}z)^{\frac{1}{2}({\rm i}B-A)}
       P_m^{(-{\rm i}B-A-\frac{1}{2}, {\rm i}B-A-\frac{1}{2})}({\rm i}z),
\end{split}
\end{equation}
with corresponding energies
\begin{equation}
\begin{split}
  & E_1 = - ({\rm i}B-m-\tfrac{1}{2})^2, \qquad E_2 = - ({\rm i}B+m+\tfrac{1}{2})^2, \\
  & E_3 = - (A+m+1)^2, \qquad E_4 = - (A-m)^2. 
\end{split}
\end{equation}
\par
%
%
The first two of these solutions being associated with a complex energy have to be rejected. Note, however, that they may be of interest for dealing with rational extensions of the $\cal PT$-symmetric Scarf II potential, obtained by replacing $B$ by ${\rm i}B$ \cite{bagchi, midya}. Considering then the remaining two solutions, we obtain $E_3 < E^{(A)}_0 = - A^2$ for all allowed parameters, while $E_4 < E^{(A)}_0 = - A^2$ imposes that $A < \frac{m}{2}$.\par
%
%
On taking Eqs.~(\ref{eq:R-P}) and (\ref{eq:R-P-parameters}) into account, $\phi_3(x)$ and $\phi_4(x)$ can be rewritten in terms of Romanovski polynomials as
\begin{equation}
\begin{split}
  \phi_3(x) & \propto (z+{\rm i})^{\frac{1}{2}(A+1+{\rm i}B)} (z-{\rm i})^{\frac{1}{2}(A+1-{\rm i}B)}
       R_m^{(2B, A+\frac{3}{2})}(z) \\
  & \propto (\cosh x)^{A+1} e^{B \arctan(\sinh x)} R_m^{(2B, A+\frac{3}{2})}(\sinh x), \qquad A>0,
\end{split}
\end{equation}
and
\begin{equation}
\begin{split}
  \phi_4(x) & \propto (z+{\rm i})^{-\frac{1}{2}(A+{\rm i}B)} (z-{\rm i})^{-\frac{1}{2}(A-{\rm i}B)}
       R_m^{(-2B, -A+\frac{1}{2})}(z) \\
  & \propto (\sech x)^A e^{-B \arctan(\sinh x)} R_m^{(-2B, -A+\frac{1}{2})}(\sinh x), \qquad 0<A<\frac{m}{2}.
\end{split}
\end{equation}
\par
%
%
At both ends of the interval $x \in (-\infty, \infty)$ or $z \in (-\infty, \infty)$, they behave as follows:
\begin{equation}
\begin{split}
  & \phi_3(-\infty) \sim \pm \infty, \qquad \phi_3(\infty) \sim +\infty, \\
  & \phi_4(-\infty) \sim \pm \infty, \qquad \phi_4(\infty) \sim +\infty, 
\end{split}
\end{equation}
where $\pm = (-1)^m$. We conclude that both are acceptable as factorization functions for building rational extensions of the Scarf II potential provided $m$ is restricted to even values. Both are also such that their inverse is normalizable so that the resulting rational extensions will have an extra bound state below the Scarf II spectrum and will therefore be of type III \cite{cq09}.\par
%
%
Since $\phi_4(x)$ corresponds to a very limited range of $A$ values (and leads to very few bound states), for simplicity's sake (see Appendix A and Ref.~\cite{footnote}) we will henceforth only consider $\phi_3(x)$, to be denoted by $\phi^{{\rm III}}_{A,B,m}(x)$ and with corresponding energy $E^{{\rm III}}_{A,m}$. It is then straightforward to find from Eqs.~(\ref{eq:partner-pot})--(\ref{eq:partner-pot-W}) that
\begin{equation}
\begin{split}
  & V^{(+)}(x) = V_{A-1,B}(x), \qquad \phi(x) = \phi^{{\rm III}}_{A-1,B,m}(x), \\
  & V_{A,B,{\rm rat}}(x) = - 2z \frac{\dot{g}^{(A,B)}_m}{g^{(A,B)}_m} - 2(1+z^2)\Biggl[
     \frac{\ddot{g}^{(A,B)}_m}{g^{(A,B)}_m} - \Biggl(\frac{\dot{g}^{(A,B)}_m}{g^{(A,B)}_m}\Biggr)^2\Biggr], \\
  & g^{(A,B)}_m(z) = R_m^{(2B, A+\frac{1}{2})}(z), \qquad m=2, 4, 6, \ldots, \qquad A>1, 
\end{split}
\end{equation}
where a dot denotes a derivative with respect to $z$.\par
%
%
The bound-state spectra of the two partners are given by
\begin{equation}
  E^{(+)}_{\nu} = - (A-1-\nu)^2, \qquad \nu=0, 1, \ldots, \nu_{\rm max}, \qquad A-2 \le \nu_{\rm max} < A-1,
\end{equation}
and
\begin{equation}
  E^{(-)}_{\nu} = - (A-1-\nu)^2, \qquad \nu=-m-1, 0, 1, \ldots, \nu_{\rm max}, \qquad A-2 \le \nu_{\rm max} < A-1,
\end{equation}
the ground state of $V^{(-)}(x)$ corresponding to $E^{(-)}_{-m-1} = E^{\rm III}_{A-1,m} = - (A+m)^2$. Note that the number of bound states does not depend on $B$ nor $m$ and is entirely determined by $A$.\par
%
%
The corresponding bound-state wavefunctions can be written as
\begin{equation}
\begin{split}
  \psi^{(+)}_{\nu}(x) & \propto (z+{\rm i})^{\frac{1}{2}(\beta-\frac{1}{2}+{\rm i}\frac{\alpha}{2})}
     (z-{\rm i})^{\frac{1}{2}(\beta-\frac{1}{2}-{\rm i}\frac{\alpha}{2})} R^{(\alpha,\beta)}_{\nu}(z) \\
  & \propto (\sech x)^{A-1} e^{-B \arctan(\sinh x)} R^{(-2B,-A+\frac{3}{2})}_{\nu}(\sinh x), \\
  & \qquad \alpha = -2B, \qquad \beta = -A+\frac{3}{2}, \qquad \nu=0, 1, \ldots, \nu_{\rm max},
\end{split}
\end{equation}
and
\begin{equation}
\begin{split}
  \psi^{(-)}_{\nu}(x) & \propto \frac{(z+{\rm i})^{\frac{1}{2}(\beta-\frac{3}{2}+{\rm i}\frac{\alpha}{2})}
     (z-{\rm i})^{\frac{1}{2}(\beta-\frac{3}{2}-{\rm i}\frac{\alpha}{2})}}{g^{(A,B)}_m(z)} y^{(A,B)}_n(z) \\
  & \propto \frac{(\sech x)^A e^{-B \arctan(\sinh x)}}{g^{(A,B)}_m(\sinh x)} y^{(A,B)}_n(\sinh x), \\
  & \qquad \alpha = -2B, \qquad \beta = -A+\frac{3}{2}, \qquad \nu=-m-1, 0, 1, \ldots, \nu_{\rm max}, \\
  & \qquad n = m+\nu+1,
\end{split}
\end{equation}
where $y^{(A,B)}_n(z)$ is an $n$th-degree polynomial in $z$.\par
%
%
{}For the ground state of $V^{(-)}(x)$, we simply have
\begin{equation}
  \psi^{(-)}_{-m-1}(x) \propto \left(\phi^{\rm III}_{A-1,B,m}(x)\right)^{-1}, \qquad y^{(A,B)}_0(z) = 1,
\end{equation}
while for the excited states, we note that $\psi^{(-)}_{\nu}(x) \propto \hat{A} \psi^{(+)}(x)$, $\nu=0$, 1,~\ldots, $\nu_{\rm max}$, with $\hat{A}$ given by Eq.~(\ref{eq:partner-factor}) or
\begin{equation}
  \hat{A} = \sqrt{1+z^2} \left(\frac{d}{dz} + \frac{\beta-\frac{3}{2}+{\rm i}\frac{\alpha}{2}}{2(z+{\rm i})} +
  \frac{\beta-\frac{3}{2}-{\rm i}\frac{\alpha}{2}}{2(z-{\rm i})} - \frac{\dot{g}^{(A,B)}_m}{g^{(A,B)}_m}\right).
\end{equation}
On applying Eq.~(\ref{eq:diff-R-bis}) both for $R^{(\alpha, \beta)}_{\nu}(z)$ and for $g^{(A,B)}_m(z) = R^{(-\alpha, -\beta+2)}_{\nu}(z)$, we arrive at the result
\begin{equation}
\begin{split}
  & y^{(A,B)}_n(z) = (\nu-m+2\beta-2) \left(- \frac{\alpha(2\beta-2)}{(2\nu+2\beta-2)(2m-2\beta+2)} + z\right)
       g^{(A,B)}_m R^{(\alpha,\beta)}_{\nu} \\
  & \quad {}+ \frac{(\nu+\beta-1+{\rm i}\frac{\alpha}{2}) (\nu+\beta-1-{\rm i}\frac{\alpha}{2})}{\nu+\beta-1}
       g^{(A,B)}_m R^{(\alpha,\beta)}_{\nu-1} \\
  & \quad {}- \frac{(m-\beta+1+{\rm i}\frac{\alpha}{2}) (m-\beta+1-{\rm i}\frac{\alpha}{2})}{m-\beta+1}
       g^{(A,B)}_{m-1} R^{(\alpha,\beta)}_{\nu}, \\
  & \quad n = m+\nu+1, \quad \nu = 0, 1, \ldots, \nu_{\rm max}.
\end{split}
\end{equation}
\par
%
%
The finite set of polynomials $\bigl\{y^{(A,B)}_n(z) \big \vert n=0, m+1, m+2, \ldots, m+1+\nu_{\rm max}\bigr\}$ of codimension $m$ provides an extension of that of (third-class) Romanovski polynomials. Their (finite) orthogonality relation
\begin{equation}
\begin{split}
  & \int_{-\infty}^{\infty} dz \frac{(1+z^2)^{\beta-2} e^{\alpha \arctan z}}{\bigl(g^{(A,B)}_m(z)\bigr)^2} 
       y^{(A,B)}_{\nu'+m+1}(z) y^{(A,B)}_{\nu+m+1}(z) = 0, \\
  & \text{if $\nu' \ne \nu$ with $\nu, \nu' \in \{-m-1, 0, 1, \ldots, \nu_{\rm max}\}$ and $- \beta - \frac{1}{2}
       \le \nu_{\rm max} < - \beta + \frac{1}{2}$},
\end{split}  \label{eq:y-ortho}
\end{equation}
directly results from the orthogonality of bound-state wavefunctions $\psi^{(-)}_{\nu}(x)$. The corresponding normalization integral, i.e., the integral in Eq.~(\ref{eq:y-ortho}) with $\nu' = \nu$, also converges for any $\nu \in \{-m-1, 0, 1, \ldots, \nu_{\rm max}\}$. Furthermore, from the Schr\"odinger equation for $V^{(-)}(x)$, it follows that the polynomials $y^{(A,B)}_{\nu+m+1}(z)$ satisfy the second-order differential equation
\begin{equation}
\begin{split}
  & \biggl\{(1+z^2) \frac{d^2}{dz^2} + \biggl[2(\beta-1)z + \alpha - 2(1+z^2) \frac{\dot{g}^{(A,B)}_m}
      {g^{(A,B)}_m}\biggr] \frac{d}{dz} \\
  & \quad {}- (\nu+m+1) (\nu-m+2\beta-2)\biggr]\biggr\} y^{(A,B)}_{m+\nu+1}(z) = 0, \quad \nu = -m-1, 0, 1, 
      \ldots, \nu_{\rm max},
\end{split}
\end{equation}
generalizing Eq.~(\ref{eq:eq-diff-R}).\par
%
%
\subsection{Rational extensions of the Rosen-Morse I potential and corresponding orthogonal polynomials}

In Ref.~\cite{cq12b}, it has been shown that the Rosen-Morse II potential has two independent polynomial-type solutions, given in Eqs.~(2.7) and (2.8), leading to rational extensions of type I, II, or III. On performing substitutions (\ref{eq:RM-formal}), we can transform them into the following two polynomial-type solutions of the Rosen-Morse I Schr\"odinger equation (\ref{eq:RM-I-eq}),
\begin{equation}
\begin{split}
  & \phi_1(x) \propto (-{\rm i})^m (1-{\rm i}z)^{-\frac{1}{2} \left(A+m+\frac{{\rm i}B}{A+m}\right)}
      (1+{\rm i}z)^{-\frac{1}{2} \left(A+m-\frac{{\rm i}B}{A+m}\right)} \\
  & \hphantom{\phi_1(x) \propto} \quad \times P_m^{\left(-A-m-\frac{{\rm i}B}{A+m}, -A-m+
      \frac{{\rm i}B}{A+m}\right)}({\rm i}z), \\
  & \phi_2(x) \propto (-{\rm i})^m (1-{\rm i}z)^{\frac{1}{2} \left(A-m-1+\frac{{\rm i}B}{A-m-1}\right)}
      (1+{\rm i}z)^{\frac{1}{2} \left(A-m-1-\frac{{\rm i}B}{A-m-1}\right)} \\
  & \hphantom{\phi_2(x) \propto} \quad \times P_m^{\left(A-m-1+\frac{{\rm i}B}{A-m-1}, A-m-1-
      \frac{{\rm i}B}{A-m-1}\right)}({\rm i}z),
\end{split}
\end{equation}
with corresponding energies
\begin{equation}
  E_1 = (A+m)^2 - \frac{B^2}{(A+m)^2}, \qquad E_2 = (A-m-1)^2 - \frac{B^2}{(A-m-1)^2}.
\end{equation}
\par
%
%
The first of these energies cannot satisfy the condition $E_1 < E^{(A,B)}_0 = A^2 - \frac{B^2}{A^2}$ for any choice of parameters, while for the second one the condition $E_2 < E^{(A,B)}_0 = A^2 - \frac{B^2}{A^2}$ imposes the restriction $A > \frac{1}{2}(m+1)$.\par
%
%
{}From Eqs.~(\ref{eq:R-P}) and (\ref{eq:R-P-parameters}), it follows that $\phi_2(x)$ can be rewritten in terms of a Romanovski polynomial as
\begin{equation}
\begin{split}
  \phi_2(x) & \propto (z+{\rm i})^{\frac{1}{2} \left(A-m-1+\frac{{\rm i}B}{A-m-1}\right)}
      (z-{\rm i})^{\frac{1}{2} \left(A-m-1-\frac{{\rm i}B}{A-m-1}\right)} R_m^{\left(\frac{2B}{A-m-1}, A-m\right)}
      (z) \\
  & \propto (\sin x)^{-(A-m-1)} e^{\left(-\frac{B}{A-m-1}z\right)} R_m^{\left(\frac{2B}{A-m-1}, A-m\right)}
      (\cot x), \quad A > \frac{1}{2}(m+1).
\end{split}
\end{equation}
\par
%
%
At both ends of the interval $x \in (0, \pi)$ or $z \in (-\infty, \infty)$, it behaves as
\begin{equation}
  \phi_2(0) \sim + \infty, \qquad \phi_2(\pi) \sim \pm \infty, 
\end{equation}
where $\pm = (-1)^m$. Such a function is therefore acceptable as a factorization function for building a rational extension of the Rosen-Morse I potential provided $m$ is restricted to even values. This extension will be of type III since the inverse of $\phi_2(x)$ is normalizable. We henceforth denote $\phi_2(x)$ by $\phi^{\rm III}_{A,B,m}(x)$ and its energy by $E^{\rm III}_{A,B,m}$.\par
%
%
Equations (\ref{eq:partner-pot})--(\ref{eq:partner-pot-W}) now lead to the following results for the two partners:
\begin{equation}
\begin{split}
  & V^{(+)}(x) = V_{A+1,B}(x), \qquad \phi(x) = \phi^{{\rm III}}_{A+1,B,m}(x), \\
  & V_{A,B,{\rm rat}}(x) = - 2(1+z^2) \Biggl\{2z\frac{\dot{g}^{(A,B)}_m}{g^{(A,B)}_m} + (1+z^2)\Biggl[
     \frac{\ddot{g}^{(A,B)}_m}{g^{(A,B)}_m} - \Biggl(\frac{\dot{g}^{(A,B)}_m}{g^{(A,B)}_m}\Biggr)^2\Biggr]
     - m\Biggr\}, \\
  & g^{(A,B)}_m(z) = R_m^{(-\alpha_{-m-1}, -\beta_{-m-1}+2)}(z), \quad \alpha_{-m-1} = - \frac{2B}{A-m},
     \quad \beta_{-m-1} = -A+m+1, \\
  & m=2, 4, 6, \ldots, \quad A>\frac{1}{2}(m-1). 
\end{split}
\end{equation}
\par
%
%
The corresponding spectra are made of an infinite number of bound states, whose energies are given by
\begin{equation}
  E^{(+)}_{\nu} = (A+1+\nu)^2 - \frac{B^2}{(A+1+\nu)^2}, \qquad \nu=0, 1, 2, \ldots,
\end{equation}
and
\begin{equation}
  E^{(-)}_{\nu} = (A+1+\nu)^2 - \frac{B^2}{(A+1+\nu)^2}, \qquad \nu=-m-1, 0, 1, 2, \ldots,
\end{equation}
respectively.\par
%
%
{}From (\ref{eq:RM-I-wf}) and (\ref{eq:RM-I-wf-bis}), the wavefunctions of $V^{(+)}(x)$ can be expressed as
\begin{equation}
\begin{split}
  \psi^{(+)}_{\nu}(x) & \propto (z+{\rm i})^{\frac{1}{2}(\beta_{\nu}-1+\frac{\rm i}{2}\alpha_{\nu})}
     (z-{\rm i})^{\frac{1}{2}(\beta_{\nu}-1-\frac{\rm i}{2}\alpha_{\nu})} 
     R^{(\alpha_{\nu},\beta_{\nu})}_{\nu}(z) \\
  & \propto (\sin x)^{A+1+\nu} e^{\frac{B}{A+1+\nu} x} R^{\left(-\frac{2B}{A+1+\nu},-A-\nu\right)}_{\nu}
     (\cot x), \\
  & \quad \alpha_{\nu} = -\frac{2B}{A+1+\nu}, \qquad \beta_{\nu} = -A-\nu, \qquad \nu=0, 1, 2, \ldots.
\end{split}
\end{equation}
Furthermore, those of $V^{(-)}(x)$ are of the form
\begin{equation}
\begin{split}
  \psi^{(-)}_{\nu}(x) & \propto \frac{(z+{\rm i})^{\frac{1}{2}(\beta_{\nu}-1+\frac{\rm i}{2}\alpha_{\nu})}
     (z-{\rm i})^{\frac{1}{2}(\beta_{\nu}-1-\frac{\rm i}{2}\alpha_{\nu})}}{g^{(A,B)}_m(z)} y^{(A,B)}_n(z) \\
  & \propto \frac{(\sin x)^{A+1+\nu} e^{\frac{B}{A+1+\nu} x}}{g^{(A,B)}_m(\cot x)} y^{(A,B)}_n(\cot x), \\
  & \quad \alpha_{\nu} = -\frac{2B}{A+1+\nu}, \qquad \beta_{\nu} = -A-\nu, \qquad \nu=-m-1, 0, 1, 2, \ldots, \\
  & \quad n = m+\nu+1,
\end{split}  \label{eq:alpha-beta}
\end{equation}
where $y^{(A,B)}_n(z)$ is an $n$th-degree polynomial in $z$.\par
%
%
{}For the ground-state wavefunction,
\begin{equation}
  \psi^{(-)}_{-m-1}(x) \propto \left(\phi^{\rm III}_{A+1,B,m}(x)\right)^{-1}, \qquad y^{(A,B)}_0(z) = 1,
\end{equation}
while for the excited states, $\psi^{(-)}_{\nu}(x) \propto \hat{A} \psi^{(+)}_{\nu}(x)$, $\nu=0$, 1, 2,~\ldots, where
\begin{equation}
  \hat{A} = - (1+z^2) \frac{d}{dz} - \frac{1}{2} \alpha_{-m-1} - (\beta_{-m-1}-1)z + (1+z^2) 
  \frac{\dot{g}^{(A,B)}_m}{g^{(A,B)}_m}.
\end{equation}
On using Eqs.~(\ref{eq:diff-R}) and (\ref{eq:diff-R-bis}) for $g^{(A,B)}_m(z) = R_m^{(-\alpha_{-m-1}, -\beta_{-m-1}+2)}(z)$ and $R^{(\alpha_{\nu}, \beta_{\nu})}_{\nu}(z)$, respectively, we arrive at the following expression for the polynomials $y^{(A,B)}_n(z)$ corresponding to $n=m+1$, $m+2$, \ldots,
\begin{equation}
\begin{split}
  & y^{(A,B)}_n(z) = - \frac{(\nu+\beta_{\nu}-1+\frac{\rm i}{2}\alpha_{\nu}) (\nu+\beta_{\nu}-1
       -\frac{\rm i}{2}\alpha_{\nu})}{\nu+\beta_{\nu}-1} g^{(A,B)}_m R^{(\alpha_{\nu}, \beta_{\nu})}_{\nu-1} \\
  & \quad + \frac{(m+1)(m-2\beta_{-m-1}+3)}{m-\beta_{-m-1}+2} g^{(A+1,B)}_{m+1} 
       R^{(\alpha_{\nu}, \beta_{\nu})}_{\nu}, \quad n=m+\nu+1, \quad \nu=0, 1, 2, \ldots. 
\end{split}
\end{equation}
For the first-excited state, in particular, we get $y^{(A,B)}_{m+1}(z) \propto g^{(A+1,B)}_{m+1}(z)$.\par
%
%
The orthogonality of the bound-state wavefunctions $\psi^{(-)}_{\nu}(x)$ with different $\nu$ values provides a generalization of the infinite set of relations among Romanovski polynomials with parameters attached to the degree, given in Eq.~(\ref{eq:infinite-ortho}). We indeed obtain
\begin{equation}
\begin{split}
  & \int_{-\infty}^{+\infty} dz\, \frac{(1+z^2)^{\frac{1}{2}(\beta_{\nu}+\beta_{\nu'})-2} 
       e^{\frac{1}{2}(\alpha_{\nu}+\alpha_{\nu'}) \arctan z}}{\left(g^{(A,B)}_m(z)\right)^2} 
       y^{(A,B)}_{m+\nu'+1}(z) y^{(A,B)}_{m+\nu+1}(z) = 0 \\
  & \quad \text{if $\nu' \ne \nu$ and $\nu, \nu' = -m-1, 0, 1, 2, \ldots$},
\end{split}  
\end{equation}
with $\alpha_{\nu}$ and $\beta_{\nu}$ defined in Eq.~(\ref{eq:alpha-beta}). Finally, it is worth pointing out the second-order differential equation satisfied by the set of polynomials $y^{(A,B)}_n(z)$,
\begin{equation}
\begin{split}
  & \biggl\{(1+z^2) \frac{d^2}{dz^2} + \biggl[2\beta_{\nu}z + \alpha_{\nu} - 2(1+z^2) \frac{\dot{g}^{(A,B)}_m}
      {g^{(A,B)}_m}\biggr] \frac{d}{dz} - (\nu+1) (2\beta_{\nu}+\nu) \\
  & \quad {}+ m (-2\beta_{-m-1}+m+1) - [2(\beta_{\nu}-\beta_{-m-1})z + \alpha_{\nu} - \alpha_{-m-1}]
      \frac{\dot{g}^{(A,B)}_m}{g^{(A,B)}_m}\biggr\} \\
  & \quad \times y^{(A,B)}_{m+\nu+1}(z) = 0, \quad \nu = -m-1, 0, 1, 2, \ldots.
\end{split}
\end{equation} 
\par
%
%
\section{CONCLUSION}

In the present paper, we have constructed rational extensions of the Scarf II and Rosen-Morse I potentials as SUSY partners of conventional potentials of the same type and shown that their bound-state wavefunctions involve some polynomials that generalize the (third-class) Romanovski polynomials (also called Romanovski/pseudo-Jacobi polynomials). In the Scarf II case, the finite orthogonality of Romanovski polynomials (in the framework defined in Sec.~II) translates into a finite orthogonality of the generalized polynomials, while in the Rosen-Morse I case, infinite sets of relations among polynomials with parameters attached to the degree are obtained.\par
%
%
The regularity of the rationally-extended potentials has been checked by taking advantage of the disconjugacy properties of second-order differential equations of Schr\"odinger type. In the absence of any direct information on the zeros of the Romanovski polynomials present in the denominators, this has illustrated the power of such properties for the construction of rational extensions once again.\par
%
%
During the construction, an extensive use has also been made of mappings between Scarf I and Scarf II potentials on one hand, and between Rosen-Morse II and Rosen-Morse I potentials on the other hand. In such a process, Jacobi polynomials have been changed into Romanovski ones. Simultaneouly, the variety of rational extensions, which included types I, II, and III for Scarf I and Rosen-Morse II potentials, has been narrowed down to only type III extensions for the resulting Scarf II and Rosen-Morse I potentials.\par
%
%
As final points, it is worth observing that these type III extensions can alternatively be derived \cite{odake13a, odake13b} in higher-order SUSYQM by using the Krein-Adler's modification \cite{krein, adler} of Crum's theorem \cite{crum} and that Romanovski polynomials have also been recently discussed \cite{natanson} in connection with Routh polynomials and the Milson potential \cite{milson}.\par
%
%
\section*{APPENDIX A: POLYNOMIAL-TYPE SOLUTIONS FOR THE SCARF I POTENTIAL}

\renewcommand{\theequation}{A.\arabic{equation}}
\setcounter{equation}{0}

By appropriate changes of variable and of function (see, e.g., Ref.~\cite{cq12b}), the Scarf I Schr\"odinger equation (\ref{eq:scarf-I-eq}) can be reduced to the hypergeometric equation, whose polynomial-type solutions, expressed in terms of Jacobi polynomials, lead to the following four polynomial-type solutions of (\ref{eq:scarf-I-eq}),
\begin{equation}
\begin{split}
  & \tilde{\phi}_1(x) \propto (1-w)^{-\frac{1}{2}(B-A)} (1+w)^{-\frac{1}{2}(B+A-1)} P_m^{(-B+A-\frac{1}{2},
        -B-A+\frac{1}{2})}(w), \\
  & \tilde{\phi}_2(x) \propto (1-w)^{\frac{1}{2}(B-A+1)} (1+w)^{\frac{1}{2}(B+A)} P_m^{(B-A+\frac{1}{2},
        B+A-\frac{1}{2})}(w), \\
  & \tilde{\phi}_3(x) \propto (1-w)^{\frac{1}{2}(B-A+1)} (1+w)^{-\frac{1}{2}(B+A-1)} P_m^{(B-A+\frac{1}{2},
        -B-A+\frac{1}{2})}(w), \\
  & \tilde{\phi}_4(x) \propto (1-w)^{-\frac{1}{2}(B-A)} (1+w)^{\frac{1}{2}(B+A)} P_m^{(-B+A-\frac{1}{2},
        B+A-\frac{1}{2})}(w), 
\end{split}  \label{eq:A1}
\end{equation}
with corresponding energies
\begin{equation}
\begin{split}
  & \tilde{E}_1 = (B-m-\tfrac{1}{2})^2, \qquad \tilde{E}_2 = (B+m+\tfrac{1}{2})^2, \\
  & \tilde{E}_3 = (A-m-1)^2, \qquad \tilde{E}_4 = (A+m)^2.
\end{split}
\end{equation}
\par
%
%
The first three of these energies are below the ground-state one, $\tilde{E}^{(A)}_0 = A^2$, provided the potential parameters are restricted by the conditions \cite{footnote}
\begin{eqnarray}
  & (1) \; &  A > m+\tfrac{1}{2}, \quad 0 < B < A-1; \nonumber\\
  & (2) \; &  A > m+\tfrac{1}{2}, \quad 0 < B < A-m-\tfrac{1}{2}; \label{eq:A3} \\ 
  & (3) \; &  A > \tfrac{1}{2}(m+1), \quad 0 < B < A-1. \nonumber
\end{eqnarray}
On the other hand, the fourth energy $\tilde{E}_4$ is always above the ground state and is actually connected with the bound-state energies (\ref{eq:scarf-E}).\par
%
%
Considering next the behaviour of the first three functions of (\ref{eq:A1}) at both ends of the interval $x \in \bigl(-Ê\frac{\pi}{2}, \frac{\pi}{2}\bigr)$ or $w \in (-1, 1)$ for the parameters given in (\ref{eq:A3}), we obtain
\begin{equation}
\begin{split}
  & \tilde{\phi}_1 \left(- \frac{\pi}{2}\right) \sim + \infty, \qquad \tilde{\phi}_1 \left(\frac{\pi}{2}\right) \sim 
       0^+, \\
  & \tilde{\phi}_2 \left(- \frac{\pi}{2}\right) \sim 0^{\pm}, \qquad \tilde{\phi}_2 \left(\frac{\pi}{2}\right) \sim 
       \pm \infty, \\
  & \tilde{\phi}_3 \left(- \frac{\pi}{2}\right) \sim + \infty, \qquad \tilde{\phi}_3 \left(\frac{\pi}{2}\right) \sim 
       \pm \infty, 
\end{split}
\end{equation}
where $\pm = (-1)^m$. As a consequence, $\tilde{\phi}_1(x)$ and $\tilde{\phi}_2(x)$ qualify as factorization functions for building rational extensions of the Scarf I potential without any further restriction on the parameters, whereas for $\tilde{\phi}_3(x)$ we have to impose that $m$ is even.\par
%
%
The three remaining polynomial-type solutions $\tilde{\phi}_1(x)$, $\tilde{\phi}_2(x)$, and $\tilde{\phi}_3(x)$ lead to rational extensions of type I, II, and III, respectively \cite{cq09}, the first two being isospectral to some conventional Scarf I potential and the third one having an extra bound state.\par
%
%
\section*{ACKNOWLEDGMENTS}

The author is indebted to an anonymous referee for drawing her attention to Ref.~\cite{kwon}. She would also like to thank G.\ Natanson for some interesting comments.\par
%
%


\begin{thebibliography}{99}
\bibitem{footnote} It is worth observing that for the first polynomial-type solutions, there also exist some exceptional cases, corresponding to very limited ranges for both parameters $A$ and $B$. Such exceptional cases, which were first signalled in Ref.~\cite{cq09} and were also encountered elsewhere \cite{cq12a}, are in general omitted for simplicity's sake.
\end{thebibliography}
\end{document}